\documentclass[sigconf,authorversion,nonacm]{acmart}
\AtBeginDocument{%
  }

\usepackage{booktabs}
\usepackage{makecell}
\usepackage{longtable}
\usepackage{listings}
\usepackage{xcolor}
\usepackage{hyperref}
\usepackage{hyperxmp}

\usepackage{array}
\usepackage{amsmath}
\usepackage{microtype}
\usepackage{graphicx}
\usepackage{graphicx}
\usepackage{float}
\usepackage{wrapfig}
\usepackage{xspace}
\usepackage{enumitem}
\setlistdepth{9}

\usepackage{lipsum}
\usepackage{todonotes}
\usepackage{tikz}
\usepackage{makecell}
\usepackage{array}
\usepackage{comment}

\graphicspath{{figures/}}

\newcommand{\sys}{\textsc{Sidecar}\xspace}

\newcommand{\kp}[1]{\textcolor{black}{#1}}
\newcommand{\sav}[1]{\textcolor{black}{#1}}
\newcommand{\mxl}[1]{\textcolor{black}{#1}}

\setcopyright{acmlicensed}
\copyrightyear{2018}
\acmYear{2018}
\acmDOI{XXXXXXX.XXXXXXX}
\acmConference[Conference acronym 'XX]{Make sure to enter the correct
  conference title from your rights confirmation email}{June 03--05,
  2018}{Woodstock, NY}
\acmISBN{978-1-4503-XXXX-X/2018/06}

\begin{document}

\title{Too Long to Watch, Too Short to Leave: Designing for Agent Micro-Waits}

\author{Kevin Pu}
\authornote{Work completed during an internship at Google DeepMind.}
\affiliation{%
  \institution{University of Toronto}
  \city{Toronto, Ontario}
  \country{Canada}
  }
\email{kevin.pu@mail.utoronto.ca}

\author{Anh Truong}
\affiliation{%
  \institution{Google DeepMind}
  \city{New York, NY}
  \country{USA}
  }
\email{anhtruong@google.com}

\author{Michael Terry}
\affiliation{%
  \institution{Google DeepMind}
  \city{Cambridge, MA}
  \country{USA}
  }
\email{michaelterry@google.com}

\author{Carrie J. Cai}
\affiliation{%
  \institution{Google DeepMind}
  \city{Mountain View, CA}
  \country{USA}
  }
\email{cjcai@google.com}

\author{Michael Xieyang Liu}
\authornotemark[2]
\affiliation{%
  \institution{Google DeepMind}
  \city{Pittsburgh, PA}
  \country{USA}
  }
\email{lxieyang@google.com}

\author{Savvas Petridis}
\authornote{Co-last authors.}
\affiliation{%
  \institution{Google DeepMind}
  \city{New York, NY}
  \country{USA}
  }
\email{petridis@google.com}

\renewcommand{\shortauthors}{Pu et al.}

\begin{abstract}

    Agentic systems increasingly carry out complex work, where a single turn routinely runs for minutes.
    This produces a micro-wait that is too long to watch and too short to leave.
    In a formative study with 12 daily agent users, we found that staying felt like wasted time, while switching away fragmented attention and forfeited the context needed to re-engage.
    We introduce \sys{}, a parallel assistant situated in the same interface as the agent, which reads the run and surfaces implicit decisions the agent made, along with task directions worth exploring.
    In a within-subjects study (N=16), \sys{} helped participants stay focused on their main task, freely workshop unfinished ideas without fear of unwanted action, and recognize under-specification in their own prompts.
    Together, these findings contribute to designs that help users stay context-aware and engaged during agent micro-waits.

\end{abstract}

\ccsdesc[500]{Do Not Use This Code~Generate the Correct Terms for Your Paper}
\ccsdesc[300]{Do Not Use This Code~Generate the Correct Terms for Your Paper}
\ccsdesc{Do Not Use This Code~Generate the Correct Terms for Your Paper}
\ccsdesc[100]{Do Not Use This Code~Generate the Correct Terms for Your Paper}

\begin{teaserfigure}
  \includegraphics[width=\textwidth]{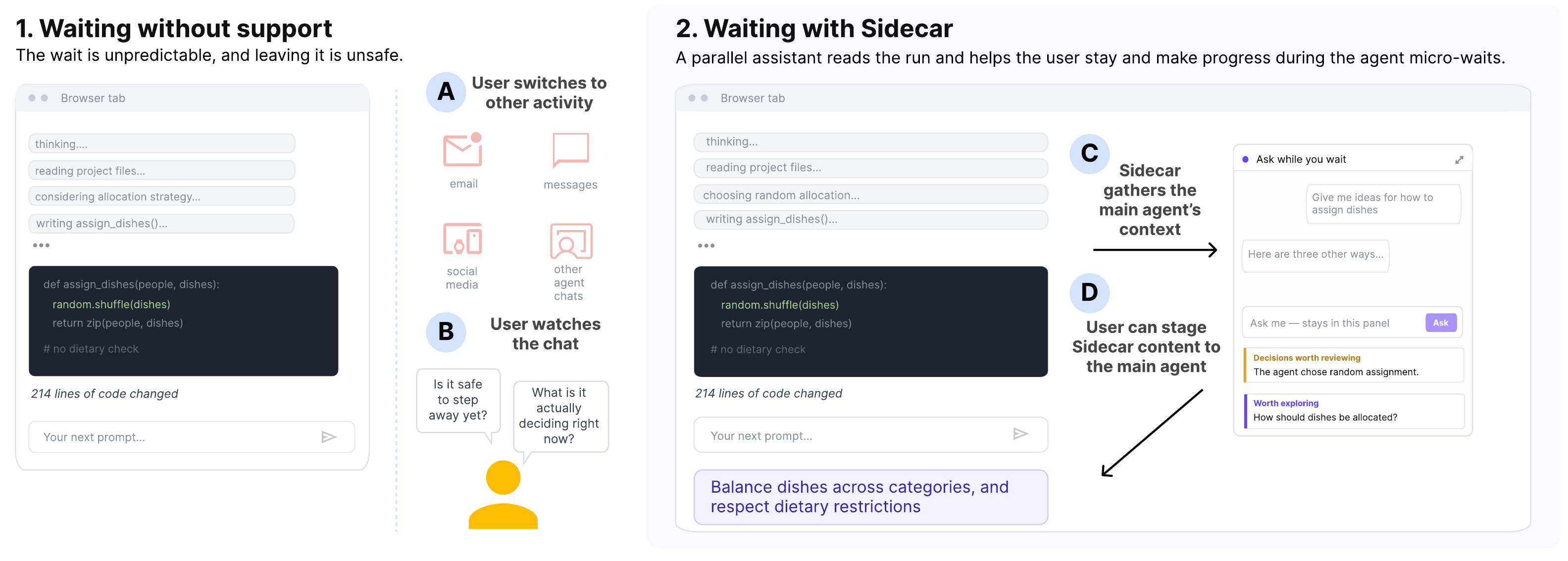}
  \caption{(Left) While waiting for agents to run, users tend to context-switch to other activities, which can fragment their attention (A), or stay and watch the agent trace, but wonder if they're wasting valuable time (B). Right: \sys{} is a side chat that sits in the same tab as the main agent chat, gathering the main chat's context (the conversation, reasoning trace, and emerging artifacts) (C). Users can chat with it similar to how they do in the main chat, but also see agent decisions to review and system-suggested directions still worth exploring. Users can ``promote" or stage content from \sys to the main chat (D). In this example, where the user is prototyping a web app for potluck dish assignments, the user asked the main chat's agent to create functionality for assigning dishes. While waiting on the agent, the user works on a related aspect of the task, asking \sys for other ideas on how to assign dishes.}
  \Description{Two-panel comparison of agent waiting with and without a
  parallel assistant.}
  \label{fig:teaser}
\end{teaserfigure}

\maketitle

\section{Introduction}
Agentic systems, with their capacity to reason over a task, call tools, and assess their own results, increasingly carry out complex work on a user’s behalf (e.g., "prototype a landing page for my app”, or “generate a short clip of a city at dusk”).
Unlike conversational assistants that reply in seconds, or predictably long-running jobs such as deep research\footnote{https://openai.com/index/introducing-deep-research/}\footnote{https://gemini.google/overview/deep-research/} that explicitly invite the user to leave and notify them on completion, many agentic systems respond in the span of a few minutes, since they often create plans, call tools, and evaluate their own responses.
\mxl{This creates a new form of interstitial time that users struggle to use productively.}
Once the user hands off the work, they're left waiting out an interval that is too long to sit and watch, too short to deeply engage in another task, and difficult to predict in length or in whether it will need their input. We call these intervals \textbf{agent micro-waits}.

With the increasing use of agentic systems, micro-waiting is becoming not only far more frequent, but also more fundamentally imbued in core knowledge work tasks. Whereas previously, waiting occurred intermittently primarily in the wild (e.g. waiting for the elevator), socially (waiting for an instant message reply), or in more specialized digital settings (e.g. waiting for code to compile), the recent emergence of agent-powered work has brought along with it a proliferation of agent micro-waits that pervade nearly every part of daily knowledge work (e.g. writing a document, generating a video, making a webpage). 
\mxl{As these waits become routine, understanding how users experience them and how systems might support users during them becomes an increasingly relevant design question.}

\mxl{To understand how users currently experience agent micro-waits, we interviewed 12 daily agent users across professional engineering, design, and research roles.}
\mxl{Because users were uncertain how long a wait would last, they faced a recurring dilemma: switching away from the page fragmented their attention and pulled them into unrelated activities, while staying on the page often felt like wasted time.}
\mxl{Users who stayed sometimes tried to plan ahead mentally, thinking about what to try next or what follow-up work might look like, but could not write down or act on those ideas without opening a separate tab and losing context of the running agent. Others found the agent's reasoning trace too dense to follow and had little else to engage with. }
\mxl{Although users wanted support for making progress during the wait, they strongly preferred that it be user-initiated and ambient.}
They did not want any interventions that might visually ``pop-up" or force them to interact with unrelated content.

Informed by these user needs, we designed \sys{}, a user interface that presents as a passive side chat that contains shared, full-context of the main agent conversation, augmented with capabilities to help users think through their main task. 
The side chat is intentionally designed to be situated in the \textit{same} tab directly adjacent to their existing agent chat, 
\mxl{so users can engage with it without context switching.}
In \sys{}, users can 1) chat in this parallel thread just as they would in their existing agent chat, but also 2) see system-suggested decisions to review and directions to explore. 

We evaluated \sys{} in a within-subjects study ($N=16$) in which participants completed tasks (prototyping a web application; generating a video) with and without \sys{}.
\mxl{
Based on post-study survey metrics and behavioral data, participants felt less attention-fragmented with \sys{} and spent less time switching away from their tasks while waiting. They felt they made more progress while the agent worked, felt more involved in what it was doing, and considered alternatives they would not have thought of on their own.
}
\mxl{In addition,}
\sys{} also helped participants 
\mxl{notice gaps in their own instructions:}
decisions they never specified and would otherwise have found only in the finished output.
At the same time, however, perceived mental workload also increased with this increased engagement; participants welcomed the added thinking about the problem itself, but spent effort interpreting suggestions shown in \sys{} and relating them back to the main agent conversation, particularly suggestions that probed too many steps ahead. These findings point to future work on when and how to engage users with minimal interpretive cost.

This paper contributes:
\begin{enumerate}
  \item An empirical characterization of agent micro-waiting from interviews with $12$ daily agent users, and the user needs people experience while waiting.
  \item \sys{}, a side chatbox situated in the same tab as the existing agent conversation, with full-context of the main agent conversation. Users can chat with \sys{} without disrupting the main agent, and also see ambient system-generated observations 
  designed to help them make progress and stay engaged with the main task.
  \item \sav{A within-subjects study ($N=16$) evaluating \sys{} against a baseline condition with no intervention, showing that \sys{} (1) reduced user tendency to switch away from their main task, and reduced their perceived attention fragmentation; (2) gave participants a channel for freely brainstorming and stashing unfinished ideas without fear of it acting on their work-in-progress; (3) surfaced the main agent's live decisions, helping participants recognize under-specification in their own prompts; and (4) worked best when suggestions stayed grounded in the agent's immediate trajectory, whereas suggestions too far ahead or too frequent increased cognitive load.}
  \item \mxl{Design implications for \sys{}-like adaptive wait-time assistance that reduces cognitive load, concurrent agent monitoring, and discussions on how agent-wait support might evolve as models become faster, more capable, and more ubiquitous.}

\end{enumerate}

\section{Related Work}
\label{sec:related}

\mxl{
\subsection{Waiting and Idle Time in Interactive Systems}
A longstanding line of HCI research has studied how people experience and cope with system delays. Miller \cite{miller_response_1968} and Card et al. \cite{card_psychology_1983} established early perceptual thresholds, with Nielsen \cite{nielsen1993usability} consolidating the widely cited guidelines that users tolerate roughly one second for uninterrupted flow and ten seconds before losing attention. Subsequent work developed interventions to manage perceived duration, including progress indicators, skeleton screens, and temporal feedback \cite{myers_importance_1985,harrison_rethinking_2007, harrison_faster_2010}. In the context of generative AI, latency has been shown to shape trust and perceived quality \cite{tan_counting_2026}, visible "thinking" indicators affect user perception \cite{cox_watching_2026}, and work on the labor illusion \cite{buell_labor_2011} found that visible signs of computation can inflate perceived output quality independent of actual results. Deliberate slowdowns have also been shown to function as productive friction in human-AI interaction, encouraging active reflection and reducing over-reliance on automation \cite{liu_how_2024, chen_exploring_2024, cox_design_2016}. While these approaches shape how the wait itself is perceived, a separate line of work asks what users might do with the time: wait-learning systems deliver educational content during delays \cite{cai_wait-learning_2015, cai_waitsuite_2017}, microproductivity tools surface lightweight tasks \cite{teevan_productivity_2016, teevan_selfsourcing_2014, cheng_break_2015, cai_chain_2016}, and active progress indicators embed useful information into loading screens \cite{hurter_active_2011}. A common property of these approaches is that the secondary content is decoupled from the primary task: the user learns vocabulary or clears a to-do item, but neither activity has any relationship to what they are waiting for. Agent micro-waits present a different situation, because the system is actively resolving the user's own task. This raises the question of whether wait-time content that is coupled to the ongoing work might serve a different function.
}

\mxl{
Whether and how to fill the wait also depends on what happens to attention during idle moments. People frequently self-interrupt, and the resulting task switches carry cognitive costs even when they are voluntary \cite{katidioti_interrupt_2016}. The standard model of interruption foregrounds resumption lag as the primary cost of leaving a task \cite{monk_effect_2008, altmann_memory_2002, borst_what_2015, adamczyk_if_2004}. Agent waits may introduce a different dynamic: because the agent may finish, stall, or request input at any moment, leaving creates two concurrent monitoring obligations rather than a clean sequential handoff. Research on restorative experiences during idle time also suggests that not all waits should be filled, as roughly 60\% of naturally occurring wait time goes to leisure activities \cite{zheng_fragmented_2024}, and microbreaks can support well-being and sustained attention \cite{kaplan_restorative_1995, rzeszotarski_inserting_2013, khot_designing_2022}. These two perspectives, that idle time can be made productive and that some idle time should be left alone, create a tension that our work engages with directly. \sys provides content during the wait, but the content is derived from the user's own task and coupled to the agent's live run. At the same time, our design treats disengagement as a valid outcome, recognizing that not every wait needs to be filled.
}

\mxl{
\subsection{Delegation and Oversight in Human-AI Interaction}
As AI agents take on increasingly complex tasks, users often communicate their intent through high-level prompts that leave many aspects of the task unspecified. LLMs frequently fill these gaps on their own, often in ways that are fragile across model and prompt changes \cite{yang_what_2026}. This pattern is amplified in agentic settings, where users compress intent into a brief prompt and bypass detailed specification \cite{sarkar_vibe_2025}, and where the gap between what users intend and what agents interpret widens as task complexity increases \cite{petridis_compass_2026}. Studies of LLM-generated data visualizations have found that when users do not specify design parameters, models make consequential choices without the user noticing \cite{ju_when_2026}. Work on scaffolding sensemaking by structuring LLM-generated content for user review \cite{liu_selenite_2024} suggests that reading model output back to users in an organized form can support comprehension, though these systems typically operate outside the time-bounded context of an agent wait.
}

\mxl{
These specification challenges sit alongside a growing body of work on overseeing and steering AI agents during execution. Classical frameworks emphasize situation awareness \cite{endsley_out---loop_1995, endsley_toward_1995}, the risk of automation complacency \cite{parasuraman_complacency_2010}, and the importance of calibrating trust and reliance \cite{lee_trust_2004, parasuraman_model_2000}. In the LLM era, researchers have explored explanations for trust calibration \cite{bansal_does_2021}, cognitive forcing functions \cite{bucinca_trust_2021}, plan-then-execute architectures \cite{he_plan-then-execute_2025}, and verification-oriented interaction patterns \cite{lee_impact_2025, sarkar_vibe_2025, cheng_it_2024, sabouri_trust_2025}. Fernandes et al. \cite{fernandes_explaining_2026} found that verbose reasoning traces can impair user performance and leave self-assessment uncalibrated. Several systems provide mid-execution oversight: WaitGPT visualizes streaming analysis code to enable monitoring and correction during LLM runs \cite{xie_waitgpt_2024}, AGDebugger supports agentic workflow debugging \cite{epperson_interactive_2025}, DirectGPT provides direct-manipulation steering \cite{masson_directgpt_2024}, and AgentGUI offers interactive interfaces for agent control \cite{zhao_agentgui_2026}. 
}

\mxl{
Among these, WaitGPT \cite{xie_waitgpt_2024} is the most closely related system to ours, visualizing streaming code to enable mid-execution correction. It operates within the agent's interface, assumes a code-legible domain, and enables direct steering. Our work differs in operating through a separate parallel channel, supporting both opaque and legible domains, and withholding actuation from the assistant. More broadly, existing oversight tools generally assume the user has time and motivation for a deliberate inspection session, whereas during agent micro-waits the user is involuntarily idle for a brief, unpredictable period. We explore what a lighter-weight form of engagement might look like in this constrained context. In doing so, we also build on work in recommendation evaluation \cite{ge_beyond_2010} that suggests the value of suggestions may extend beyond adoption rate, including their role in prompting the user's own reflection.
}

\mxl{
\subsection{Proactive Assistance and Peripheral Information}
Research on mixed-initiative systems has long studied the trade-off between proactive system initiative and user autonomy \cite{ horvitz_principles_1999, amershi_guidelines_2019}. Proactive assistance can be valuable when well-timed, but it risks interrupting the user's own reasoning. Work on notification timing and receptivity has identified natural breakpoints where interruptions are least costly \cite{iqbal_understanding_2007, iqbal_effects_2008}, and studies of proactive programming support have examined how AI tools can surface suggestions without derailing the developer's train of thought \cite{pu_assistance_2025}. Across this literature, the cost of proactive help generally scales with how busy the user is at the moment of delivery. Agent micro-waits present a distinct case: because the user is already waiting, the existence of this interstitial time may reduce the cost of surfacing information relative to typical proactive-assistance scenarios. Our work examines what happens when a system takes advantage of this micro-moment to surface task-relevant content.
}

\mxl{
The question of how to present such information connects to a broader design tradition around parallel and peripheral interfaces. Side-panel and parallel-channel interfaces have become common in AI-assisted tools, from code completion sidebars \cite{barke_grounded_2023} to browser-based AI assistants \cite{jiang2026orca} and side-by-side model evaluation interfaces \cite{kahng_llm_2024,kahng_llm_2024-1}, though these tools typically embed the assistant within the primary application context. A related line of research, rooted in Weiser and Brown's notion of calm technology \cite{weiser_coming_1997}, emphasizes placing information at the periphery of attention so that it is available without demanding focus, as exemplified by peripheral awareness displays \cite{matejka_ambient_2011} and ambient information systems \cite{mankoff_heuristic_2003}. \sys{} draws on both traditions, providing a parallel channel for information delivery that is designed to minimize intrusiveness. It applies prior peripheral display strategies, such as being easy-to-initiate (sitting on the same page as the main chat), easy-to-dismiss (requires no user action to dismiss), and is a static display with changing content rather than a pop-up that appears and disappears. Where it differs from prior systems is that its content is derived from a live agent run, and is intended to help support and further their progress on the user's agent-powered task. %
}
\section{Formative Study}
\label{sec:formative}

To ground our design in current practice, we first interviewed people who delegate multi-step work to AI agents daily, asking what they actually do during agent waits, what makes those waits costly, and what support they would accept. Then over a few weeks, we ran a second phase of follow-up sessions in which participants worked with in-progress prototypes of \sys{} in 30-minute sessions, using their feedback to refine the design into the final version we evaluate in Section~\ref{sec:system}.

\subsection{Method}
For the interviews, we recruited 12 participants (S1--S12) at a large technology company, screening for regular use of at least one agentic tool with waits ranging from 15 seconds to several minutes; eleven reported using such tools multiple times a day.
The sample spans four software engineers, four design roles, and four researchers, and covers coding agents, deep-research tools, conversational assistants, and image and video generation. Sessions ran approximately 30 minutes over video call and were recorded and transcribed with consent.
We used critical-incident recall~\cite{flanagan1954critical}, asking participants to walk through their two or three most recent real waits rather than describe typical behavior, then probed transitions away from and back to the agent and elicited desired support.
We conducted inductive open coding followed by reflexive thematic clustering~\cite{braun2006using, braun2021thematic}, yielding 13 themes over \kp{60} codes; below we report the five findings that most directly motivated \sys{}.

\subsection{Findings}
\label{sec:formative-finding}
Participants were uncertain of how long agentic systems might take for a task, and because of this, they either switched away from the page while waiting, frequently resulting in lost focus and attention fragmentation, or they stayed on the page waiting, but felt they lost valuable time in doing so.
Participants described that it was very difficult for them to predict how long an agentic wait would last, noting that the agentic system might complete the task \textit{``within five seconds, or like an hour''} (S6).
Consequently, participants felt unable to determine what to do during that wait time.
Faced with this uncertainty, most participants (10 of 12) noted that they often watch the agent, to catch any errors or wrong assumptions it might be making, but this often felt like wasted time and effort.
S1 described a time where they caught an agent \textit{``looking at the wrong folder''} and had to stop the run.
And while they sometimes successfully catch a bug or two, participants mostly described agent traces as illegible, with S12 summarizing their experience as \textit{``I don't know what it's doing.''}
Meanwhile, participants who switched away during the wait instead found that their attention became fragmented, landing somewhere between rest and work without qualifying as either.
Eleven of twelve reported switching to another work task, but seven reported phone or social media use, describing these latter activities as \textit{``not a good use of time''} (S2).
Switching to another task would often absorb their attention, taking them away from their main task longer than they intended and leading to a loss in focus: \textit{``the worst is if I open Twitter\ldots I'll suck more time into it than I supposed to''} (S6).

When participants stayed on the page waiting, some naturally tried to make forward progress in their heads, thinking about what else to try or what follow-up work might look like, but could not externalize these thoughts without leaving the page.
Some responded by opening a second tab or chat window, either to jot down notes or to ask questions and plan subsequent features; S6 described often having \textit{``multiple questions about the thing that's running.''}
Switching away, however, introduced two costs.
First, opening a brand new tab meant losing context of the job they were already running.
Second, the new tab or chat lacked the context of the previous run, so participants lost the ability to ask questions grounded in that prior work, sometimes initiating another agentic run just to gather context all over again.
This points to a \textbf{need for a surface, situated in the same space as the agent run itself rather than a separate tab, that lets users brainstorm and ask questions with full context of what the agent has already done}.

Other participants who wanted to be productive during the wait were not sure what else to think about aside from staring at the agent's reasoning traces.
Trying to make sense of the trace ultimately was not a fruitful task: it took up their attention, and it was hard to actually pull out key decisions or errors from the trace itself.
Even close attention did not reliably pay off, S6 reported missing a run's completion despite watching for it, saying this happened "\textit{all the time}."
For participants without a technical background, this problem was compounded further: S12 described being unable to evaluate what the agent was doing at all, defaulting to ``\textit{I basically trust it}.''
Even when participants did manage to identify a decision or misstep, it was still difficult to go from that decision to thinking about alternatives, with friction involved in moving to another tab to brainstorm.
It was difficult to keep up with the full context of what the agent had done and to think about logical next steps or branching points, underscoring a \textbf{need to help users break out of their one-track mind and envision next steps}.

Finally, although participants desired support for making progress while waiting, they strongly preferred that these activities be designed to be ambient and user-initiated.
Participants wanted productivity support to be ambient and easy to dismiss, whereas if it were featured prominently and demanded attention, the reaction was resistance: S8 anticipated \textit{``I'm going to immediately have this reaction of like don't tell me what to do.''}
Suggestions should be few and should expand on or stay closely aligned with what the user is currently doing, rather than occupying their time or introducing too big a divergence.
S8 envisioned support that helped \textit{``color around what you're doing as opposed to here's stuff to occupy your time.''}
Participants also suggested that productivity support did not need to appear all the time.
While S5 noted they would appreciate support they also qualified by saying: \textit{``not every window of time needs to be filled with a system.''}
Above all, participants wanted to keep their own initiative and only get support when they wanted or needed it: S6 observed that \textit{``it's different when your mind naturally goes to the other task... easier to transition compared to if you were told.''}
Together, these findings point to \textbf{a design for support that is user-initiated, ambient, and easy to ignore}.

\subsection{Design Goals}

From these findings, we established the following design goals for \sys{} (\S\ref{sec:system}):

\begin{enumerate}
    \item \textbf{Give users a way to make progress on their task without physically switching away from it.} Participants who tried to externalize their thinking while waiting, asking questions, or planning next steps, had to leave the run to do so, at the cost of losing sight of the job in progress and losing the context needed to ask informed questions. \sys{} should offer a surface co-located with the run itself, so that users can think, write, and ask questions without ever navigating away.
    \item \textbf{Help users digest the agent's reasoning trace and break out of tunnel vision.} Participants who stayed and watched the trace directly found it illegible and unproductive to parse on their own, struggling to extract key decisions and even more so with considering alternatives and follow-up work. \sys{} should surface key decisions from the trace and offer next step suggestions to help users break out of their one-track mind.
    \item \textbf{Keep support ambient, user-initiated, and easy to dismiss.} \sys{} should present support ambiently, through a persistent but unobtrusive presence rather than intrusive pop-ups, so users can engage with it when it matches their needs and disregard it otherwise.

\end{enumerate}

\section{\sys{}: User Interface and System Design}
\label{sec:system}

\begin{figure*}[t]
  \centering
  \includegraphics[width=\textwidth]{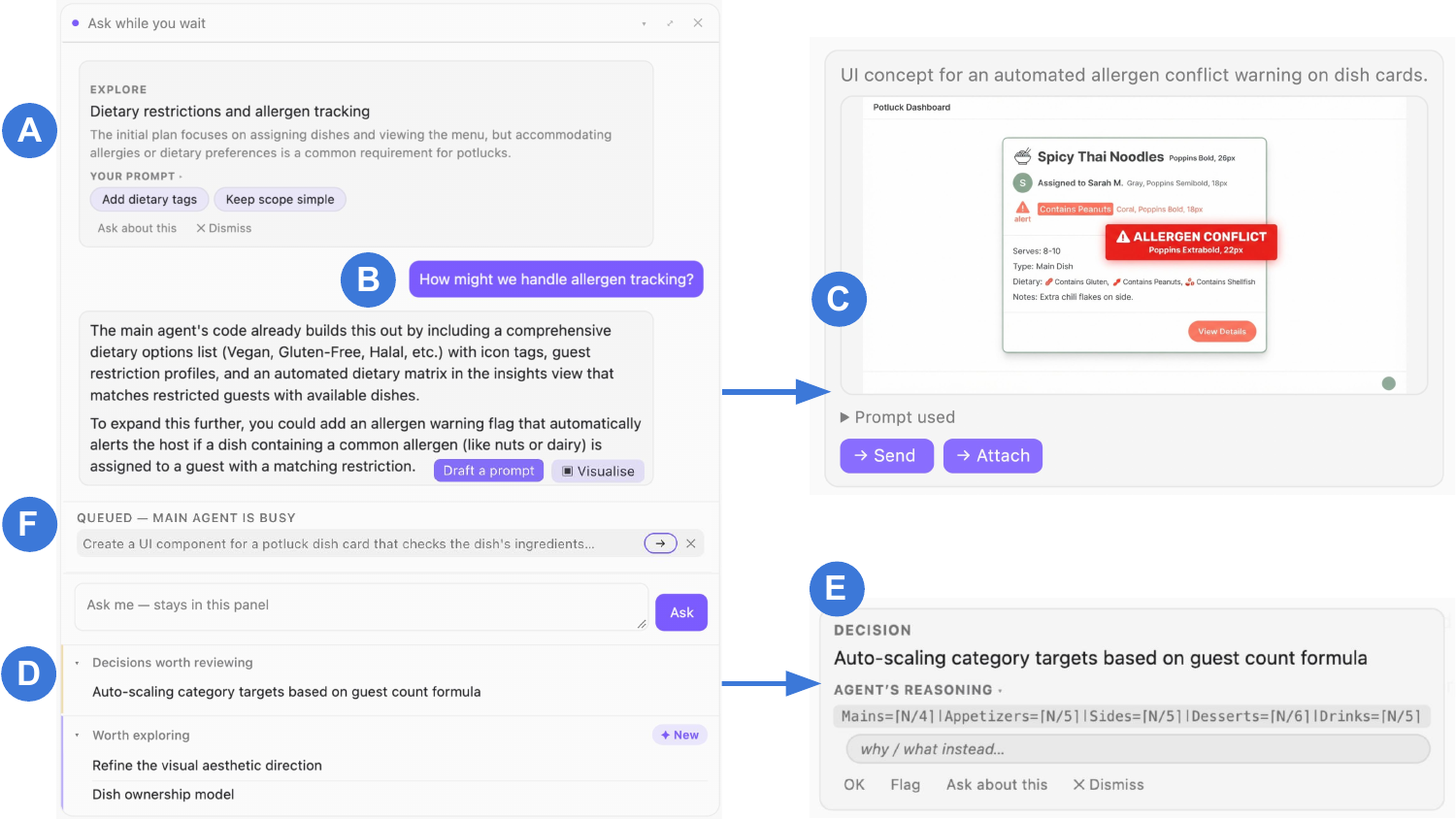}
  \caption{\sys{} interface, which consists of a chat (B) with which users can brainstorm and ask about the main chat agent's work. Underneath this chat, are extracted decisions to review from the main agent's work and suggested items to explore (D). Clicking on them reveals a card with either follow-up prompt options to explore (A), or the main agent's made decision grounded in context (i.e. code) (E). Users can also generate queued prompts for the main agent (F) or images for a given message and attach them as context or reference in the main chat (C).}
  \Description{}
  \label{fig:user-scenario}
\end{figure*}

\sav{From these design goals, we developed \sys{}, a browser extension that presents as a side chat that contains shared, full-context of the main agent conversation, augmented with capabilities to help users think through their main task.
In the following section, we illustrate \sys{} through a usage scenario that reflects the types of tasks participants encountered in our evaluation study (Sec.\ref{sec:tasks}). While the scenario is constructed, the task domains, agent platforms, and wait patterns it depicts are representative of those used in the study. Following this usage scenario, we discuss how each design goal was realized in the interface.}

\subsection{Usage Scenario}

Maya is a product manager building an exploratory prototype of a tool that helps users organize potlucks.
She begins by entering a broad prompt, describing that the web application should let users assign dishes to guests and be able to view the menu.
\mxl{The agent begins working, and its thinking traces start appearing as it generates server code, an HTML skeleton, and supporting files.}
\mxl{Normally, she might spend this fragmented time to check her phone or email while she waits, or try to follow along by reading the agent's reasoning trace.}
But this time, she notices a suggestion in \sys{} under ``Worth Exploring'', entitled ``dietary restrictions and allergen tracking'' (initially situated in Fig.\ref{fig:user-scenario}.D, then Maya promoted it to A).
\mxl{She had not yet considered how the application might handle dietary restrictions, and seeing that the agent is still mid-run, she decides to explore the idea while she waits.}
She clicks on the suggestion, which reveals a small card in the \sys{} chat (Fig.\ref{fig:user-scenario}.A), where she can see \mxl{several concrete approaches, including adding dietary tags to guest profiles.}

Interested in exploring this further, Maya types \mxl{a follow-up question into \sys{}'s chat (Fig.\ref{fig:user-scenario}.B): ``How might we handle allergen tracking?''}
\mxl{Because \sys{} has access to the main agent's full context, it can ground its response in the code already written. It explains that the agent has already implemented dietary tags for guests, and suggests adding a warning flag that alerts users when a guest is assigned a dish containing one of their allergens. (Fig.\ref{fig:user-scenario}.B)}
Curious about what \mxl{the} flag might look like, Maya presses a ``Visualize'' button on the \sys{} message, which \mxl{generates a visual mockup} of the warning (Fig.\ref{fig:user-scenario}.C).
\mxl{She likes the design and presses `attach' to stage the image in the main agent's composer, ready to include in her next prompt.}
She can also press ``Draft a prompt'' (Fig.\ref{fig:user-scenario}.B), which will stage a follow-up message in \sys{}'s queue for implementing the warning flag (Fig.\ref{fig:user-scenario}.F). This queued message can be promoted to the main agent's composer and be ready to fire once the wait is over.

Satisfied with this next step on adding warning messages for allergies, Maya glances back at the main \mxl{agent} chat to 
\mxl{see it is still working.}
She can see that \sys{} ambiently offered a few more suggestions 
\mxl{about the application's visual aesthetic}
but she is not interested in that at the moment and decides to 
\mxl{look past it}.
\mxl{What catches her eye is a line under ``Decisions worth reviewing'' (Fig.\ref{fig:user-scenario}.D): the agent has chosen a formula for calculating how many dishes to prepare based on guest count.}
\mxl{Clicking on it expands the card in the chat, which reveals the attached code (Fig.\ref{fig:user-scenario}.E), showing the formula the agent wrote.}
She thought it might be useful to update this formula in the future to account for varying appetites and preferences among her guests, and she considers engaging the \sys chat with this thought.
Overall, with \sys{}, 
\mxl{while the agent worked, Maya stayed with her task, explored a direction she had not considered, ignored suggestions that did not interest her, and prepared follow-up instructions for the agent's next turn.}

\subsection{Making Progress Without Switching Away (DG1)}
Our formative study found that participants who tried to plan ahead or develop follow-up ideas during waits had to leave the run to do so, opening a second tab or chat window and losing both sight of the agent's work and the context they needed to ask informed questions (Sec.\ref{sec:formative-finding}). We address this by providing a co-located surface, the \sys, that sits alongside the agent's chat in the same tab (Fig.\ref{fig:teaser}. Right) and gives users a place to think, write, and ask questions without navigating away. The \sys is always visually present on screen, but can be optionally minimized when the user is not engaging with it, preserving its content for later without occupying screen space. Users can also drag to move the \sys to anywhere else on the screen: this is essential because different agentic platforms have unused whitespace in different areas of the page. They can also dock \sys to the main chat itself, right above the main composer. 

\sys's system logic activates automatically when a presumed wait begins, detected from the browser's network behavior: a wait opens when a request sends the user prompt. It continuously detects the returning streaming response and detects wait-close when that stream ends. 
Within a wait, a context bundle assembles a description of what is happening, organized into independently requestable tiers: the conversation so far, the prompt in flight, the user's unsent text in the main agent composer, the live reasoning trace, the main agent's editor and canvas contents, code captured from the response stream, rendered previews, and environment facts such as elapsed time. 
This context bundle enables both the cards described in Section \ref{sec:sidecar-features} and the user's ability to ask grounded questions without leaving.

\mxl{
Because the \sys chat has access to the full context bundle, the user can ask questions grounded in what the agent has actually done, develop a direction through further messages, or request visualizations of ideas that would otherwise be unavailable until the run completes (Fig.\ref{fig:user-scenario}.C).
The \sys is augmented with additional system-suggested features, \textit{Decision} cards and \textit{Explore} cards (more details in Section \ref{sec:sidecar-features}). When something in these cards catches the user's attention, the user can open it into the chat (Fig.\ref{fig:user-scenario}.A).
From there, users can use the action button to stage content into the \sys's prompt queue with one click (Fig.\ref{fig:user-scenario}.F). Promotion to the main agent composer requires a user action. 
Users can engage as little or as much as they choose, from glancing at a card to eventually sending a message to the agent.
}

\mxl{
\subsection{Digesting the Trace and Breaking Out of Tunnel Vision (DG2)}
\label{sec:sidecar-features}
Participants who stayed on the page and watched the agent's reasoning trace found it difficult to parse on their own, struggling to extract key decisions and even more so to consider alternatives or logical next steps (Sec.\ref{sec:formative-finding}). \sys{} addresses this by running two analysis processes over the live context bundle and surfacing their output as structured cards in the \sys, rather than asking users to read the raw trace.
}

\mxl{
\sys surfaces two types of suggestions (Fig.\ref{fig:user-scenario}.B), which can be promoted into the side chat as interactive cards (Fig.\ref{fig:user-scenario}.A, E). 
\textit{Decision} cards name choices the agent has implicitly made that the user's prompt did not explicitly specify, such as picking a particular layout, choosing a formula for calculating quantities, or interpreting an ambiguous instruction in a specific way. 
These let the user consider whether the agent's choice aligns with what they had in mind. 
\textit{Explore} cards name paths the task could take that the user has not yet committed to, such as adding dietary restriction tracking to a potluck organizer. 
These give the user something concrete to react to during the wait. 
The two types are visually distinguished because they invite different responses: a decision card asks whether the user intended this choice, while an explore card asks whether a new avenue is worth exploring.
Each card also carries the evidence it was grounded in any part of the context bundle with snippets to indicate provenance, reachable behind a citation control (Fig.\ref{fig:user-scenario}.A, E).
Some cards cite the agent's reasoning trace, streamed code, or a rendered artifact (e.g. images, frames from a generated video), while others draw from the user's original prompt, and the system tracks which evidence tiers were available so the user can judge how much weight to give each card.
}

\mxl{
Two capture mechanisms go beyond what is visible on screen to support these cards. 
On coding surfaces, agents write long files behind the scene, not fully displaying the codebase to the user as it builds up, so \sys{} reconstructs the full code from the intercepted response stream as it arrives. 
On generation surfaces such as image and video, \sys{} samples visual frames when a generation artifact mounts in the page. 
The decision extractor requires a minimum evidence threshold before running, which reduces the risk of generating cards with no meaningful grounding. Before that threshold is met, \sys{} issues preflight suggestions drawn from the user's prompt, so the user sees content from the first seconds of a wait, with the cards becoming more grounded as the run progresses.
}

\mxl{
\subsection{Keeping Support Non-intrusive and User-Initiated (DG3)}
\label{sec:nonintrusive-features}
Participants in our formative study strongly preferred that wait-time support be non-intrusive and easy to dismiss (Sec.\ref{sec:formative-finding}). Importantly, we had initially considered a different design that triggered entire cards of content to appear only during wait times and disappear if not interacted with. However, that was far too visually salient and immediately grabbed users' attention. Furthermore, because users' thought processes varied widely while waiting, surfacing unwelcome content in a visually salient way would risk interrupting the user. Thus, rather than having Decision and Explore cards appear and disappear saliently, we instead provided users with an always-present side chat box (giving them the agency to type anything into the chat), and reduced Decision and Explore cards to much smaller sections below the side chat. Rather than having those cards appear and disappear in entirety, we instead kept those cards always present in \sys, but had the \textit{text} within the cards appear and disappear instead, which was less visually salient. 
}

Three signals shape when and how much content \sys surfaces, so that its pacing matches the situation. Elapsed wait time sets an emission budget that increases as the wait goes on, meaning a three-minute video generation produces more suggestions than a thirty-second coding turn. Initially, we had explored using the LLM itself to predict wait duration from early-run signals (e.g. the prompt and environment context), but because the predictions did not reach sufficient accuracy, tended to over-estimate wait time, and were heavily dependent on models and platforms, we chose to rely on measured elapsed time instead. 

Second, the delta of new evidence from the agent also plays a role, since more task progress indicates the agent has made more decisions, and more options can be open to explore. 
A third signal, inferred user stance via a lightweight LLM-evaluator based on the user's prompt, tunes the frequencies between explore cards and decision cards depending on whether the user appears to be exploring broadly or narrowing in and validating agent decisions. 

\mxl{
The \sys suppresses new offers while the user is typing in the agent's composer, avoiding competition with active composition. The suggestion content is drawn from the work already underway, and each card can be read with a glance or left alone, with no pop-ups or required responses. As the main agent's context evidence accumulates, \sys evaluates suggestions that the user left alone and retires them if they become stale. We showed the decisions and explorations directly in the UI (rather than requiring a user to click to open them), in order to make it easy for users to glance over and be able to decide whether to engage further or ignore it, without incurring the physical interruption of having to click. 
}

\subsection{Implementation}

\sys{} is implemented as a Chrome extension using 62 JavaScript modules\footnote{We implemented \sys{} with the help of agentic coding tools}. The extension separates network interception and context assembly from inference and UI rendering, with a service worker proxying LLM model calls. Platform-specific differences, such as scraping, composer autofill, and frame extraction, are confined to adapter modules, allowing \sys{} to operate across the two platforms used in our study: AI Studio (\texttt{veo-3.1}) for video generation and Antigravity 2.0 (\texttt{gemini-3.7-flash}) for prototyping. Suggestion generation and situation inference are powered by \texttt{gemini-3.5-flash-lite}, chosen for its low latency at the quality level needed for glanceable cards. Image generation in \sys uses \texttt{gemini-3.1-flash-lite-image} to support the Visualize feature. Across both, \sys{}'s generation took a median of 1.46 seconds for text and 3.99 seconds for images during the user study. Wait open and close events are logged unconditionally, independent of whether the \sys{} interface is active, so the log captures waits the user spent entirely outside the interface. The interaction design is independent of the specific models used, and we expect the approach to transfer as model capabilities evolve. \kp{\sys{}'s LLM system prompts are included in the supplementary materials.}

\section{User Study Evaluation}

We conducted a within-subjects user study with two conditions: a \sys condition and a baseline condition. The \sys condition shows the dock UI, with the decisions and exploration suggestions attached. The \sys UI can be dragged, resized, or minimized by the user. The baseline condition is identical to the experimental condition, but with the \sys UI removed entirely, so the user sees the default platform UI (the main agent chat conversation, and anything else that appears by default on the platform). This is to provide a control condition where the user experiences the exact same agent wait to isolate the effects of \sys. Our evaluation was guided by three research questions:
\begin{itemize}
  \item \textbf{RQ1:} How does \sys affect attention allocation and focus, and at what cost in mental effort?
  \item \textbf{RQ2:} How does \sys affect user ability to make forward progress on the task while waiting?
  \item \textbf{RQ3:} How do participants use \sys in their thought process and agent navigation? What patterns of use and strategies arise?
\end{itemize}

\subsection{Participants}
We recruited 16 participants from a large technology company (V1--V8 for video tasks, P1--P8 for prototyping tasks, see below for task rationale), screening for regular use of at least one agentic tool with waits of roughly 15 seconds to several minutes and prior experience in the task domain. The eight prototyping participants were four software engineers, two UX
engineers, and two UX designers. While waiting, seven noted switching to another work task and six noted watching the agent's progress.
The eight video participants were five design-adjacent roles, two product managers, and one deployment engineer; All eight noted switching to another work task while waiting, and four watching the agent's progress. No participant had prior exposure to \sys. Each session lasted approximately 60 minutes and participants were provided with \$25 for participating.

\subsection{Tasks}
\label{sec:tasks}
Participants completed two open-ended agentic tasks, one with \sys and one with the baseline UI (order counterbalanced). Each participant completed both conditions in a single domain: either two \emph{video generation} tasks (8 participants) or two \emph{web prototyping} tasks (8 participants). The two \emph{video generation} tasks asked participants to produce a set of three short clips that read as belonging together: a promotion for a restaurant opening, and a travel promotion for a city of the participant's choosing. The two \emph{prototyping} tasks asked participants to build a small working web tool: a group potluck dish-assignment tool, and an interactive browser game with progression. \kp{Tasks were conducted on commercial agent platforms: AI Studio running \texttt{veo-3.1} for video generation, and Antigravity 2.0 running \texttt{gemini-3.7-flash with high thinking} for prototyping.}

We included two task domains because process legibility differs sharply between them. Prototyping exposes a live reasoning trace and an incrementally updating artifact, so \sys reasons over the run as it unfolds; video generation exposes much less thinking traces mid-run and only discrete video outputs. This lets us examine whether coupled wait support depends on a legible process or holds where the wait is less transparent.
Every task description is deliberately under-specified so that the user will need to reason about and plan how they want to achieve the task.

\subsection{Procedure}
Condition and task instance orders were counterbalanced using a balanced Latin square within each task domain and across 16 participants.
At the start of the study, participants were encouraged to react to the wait as naturally as possible, including doing any other activities they would normally do while waiting, so as long as they kept the task agent running without physically closing that page.
Before each task, participants watched a tutorial of the system they are using, then submitted a first prompt based on the task description. 
Participants worked on the task for around $15$ minutes while using a think-aloud protocol. The facilitator remained silent except to resolve tooling failures. 

After both tasks, participants completed a Likert survey assessing understanding of the main task agent, exploration and task understanding, engagement, and perceived workload in both conditions. They then completed a semi-structured exit interview comparing their experiences across conditions.

\subsection{Data Collection and Analysis}
We collected three data streams: (1)~post-task surveys (7-point Likert items on understanding, exploration, and engagement, plus selected NASA-TLX workload items~\cite{HART1988139}); (2)~system interaction logs recording per-wait duration, attention state over time (main agent tab, or away), prompts submitted, and in the \sys condition, which suggestions were surfaced, which were expanded, and whether an expanded suggestion produced a main-agent prompt; and (3)~semi-structured exit interviews. The sessions were conducted over video-conferencing and were screen- and audio-recorded for analysis. Participants shared their screen on the main task agent's browser tab with \sys, and no other tabs to protect privacy if they choose to task-switch.

For survey data, we report descriptive statistics with two-sided Wilcoxon signed-rank tests with Benjamini–Hochberg false discovery correction at q = .05; we report BH-adjusted significance throughout and note uncorrected results only where stated.
For system logs, we compared conditions and task domains on attention allocation during waits, wait counts and durations, and prompt counts.
For interview data, we conducted inductive thematic analysis~\cite{braun2006using, braun2021thematic}, open-coding all transcripts and iteratively grouping codes into themes grounded in participant quotes.

\section{Results}
\label{sec:results}

We report findings from $N{=}16$ participants (V1--V8 video generation, P1--P8 prototyping),  119 logged waits (56 in baseline and 63 in \sys condition), and 16 paired survey responses.

\begin{figure*}[t]
  \centering
  \includegraphics[width=\textwidth]{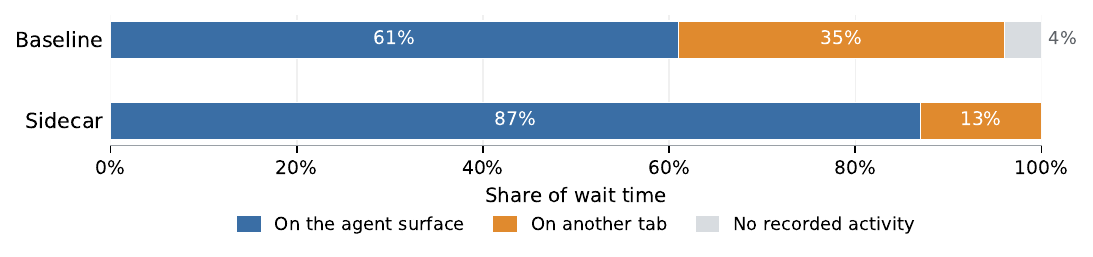}
  \caption{Where did the user spend time during the wait.
    Time on the agent tab include interacting with the agent page in the Baseline condition and also with \sys in the experimental condition. 
    \emph{No recorded activity} is wait time when the participant
    switched out of the browser window altogether with no interaction logged on any browser tab (e.g. switched away from the browser to a different application like an IDE). Time spent on other browser tabs was 35\% in the baseline, and 13\% with \sys.}
  \label{fig:timealloc}
\end{figure*}

\subsection{Wait-time behaviors}
\label{sec:whatdid}
In this section, we describe behaviors users exhibited while waiting for agents to execute, based on user think-aloud comments, exit interviews, and interaction logs. Note that we do not record the contents of off-study browser tabs for privacy reasons, so what participants did while they were on other tabs or windows comes from their think-aloud comments and interviews alone.

Wait durations did not differ significantly between baseline ($Mdn=76$ seconds) and \sys{} ($Mdn=72$ seconds). But based on interaction logs, participants spent more time switched away from the working agent's tab in the baseline ($35\%$ of wait-time) compared to the \sys condition ($13\%$) while waiting for agents to run (Fig.\ref{fig:timealloc}). They also left the tab more frequently in the baseline condition (26 of 56 waits) vs. \sys condition (7 of 63 waits).

In the baseline condition, participants read agent reasoning traces, reviewed agent-generated code, brainstormed ideas, or sat idle when staying on the page. When they switched away from the page, participants described a wide range of side tasks: V2 listed texting their partner, shopping for a gift, checking work messages, catching up on email, and reviewing their calendar; V3 spent a wait \textit{``responding to a bunch of questions my manager had''}; V8 browsed lunch menus, chat, and email. P3 described staying in the tab but reaching for their phone specifically so they could keep the agent visible on screen for monitoring. Many participants in the web prototyping task wanted to do other programming-related tasks while waiting, such as review code pull requests, or work on other agentic programming tasks in other agent chat tabs. Two users in the video task (V1, V8) reported making forward progress on their tasks while in another tab, by opening a new agent tab to brainstorm a restaurant name, or a logo for their video. Notably, they picked an atomic sub-task (restaurant names, logo designs) that did not require significant context to be carried over from the main task to their new agent chat.

In the \sys condition, participants also read agent traces and brainstormed ideas while waiting, but spent more of their time using the side chat to make forward progress on their task, and less time spent on other tabs. Participants used the side chat in different ways for the video-creation task versus the web prototyping task.
In the video-creation task, participants used the side chat to draft new material to seed their main run, to explore diverging ideas from the main branch in parallel, or to work on a different dimension of the task while the main task was running. For example, users asked the side chat to draft image key frames (e.g. a scene, a logo) or write prompts (to seed the next video-generation run in the main chat); visualize the video in a variety of settings like daytime, golden hour, and nighttime (to explore divergent ideas from the main branch); or brainstorm the personality of the main character (to work on a different dimension from the main task). 

In the prototyping task, participants used the chat predominantly to clarify the agent's decisions, or to develop strategies to use in the main chat. For example, P4 saw an extracted decision on the main agent's approach for processing audio, and asked \sys to explain this decision: \textit{``why is this API used?''} While building a snake-like game, P1 asked in the side chat to brainstorm ways to simplify the game's interactions, and asked follow-up questions based on its ideas. Notably, as these questions tended to be strategic rather than directly code-generating, \sys enabled users to bifurcate themselves, working on both types of data in parallel without ``polluting" the other chat.

Participants sent a total of 54 messages in \sys{}'s chat, with 14 of 16 participants typing in it at least once.
At the same time, participants engaged quite a bit with \sys{}'s surfaced decisions and suggestions.
\sys{} provided 58 exploration suggestions and 54 extracted decisions from the main agent's thinking traces and artifacts.
Of 112 suggestions surfaced, 45 (40\%) were expanded as cards. Twenty-eight were staged into the main-agent composer, of which 14 followed a suggestion option-click and 14 came from text the participant composed in \sys. Half of staged text (14/28) was ultimately sent as a main agent turn without further user prompt editing. Only 5 (4.5\%) suggestions were explicitly dismissed by the user.

\begin{figure*}[t]
  \centering
  \includegraphics[width=\textwidth]{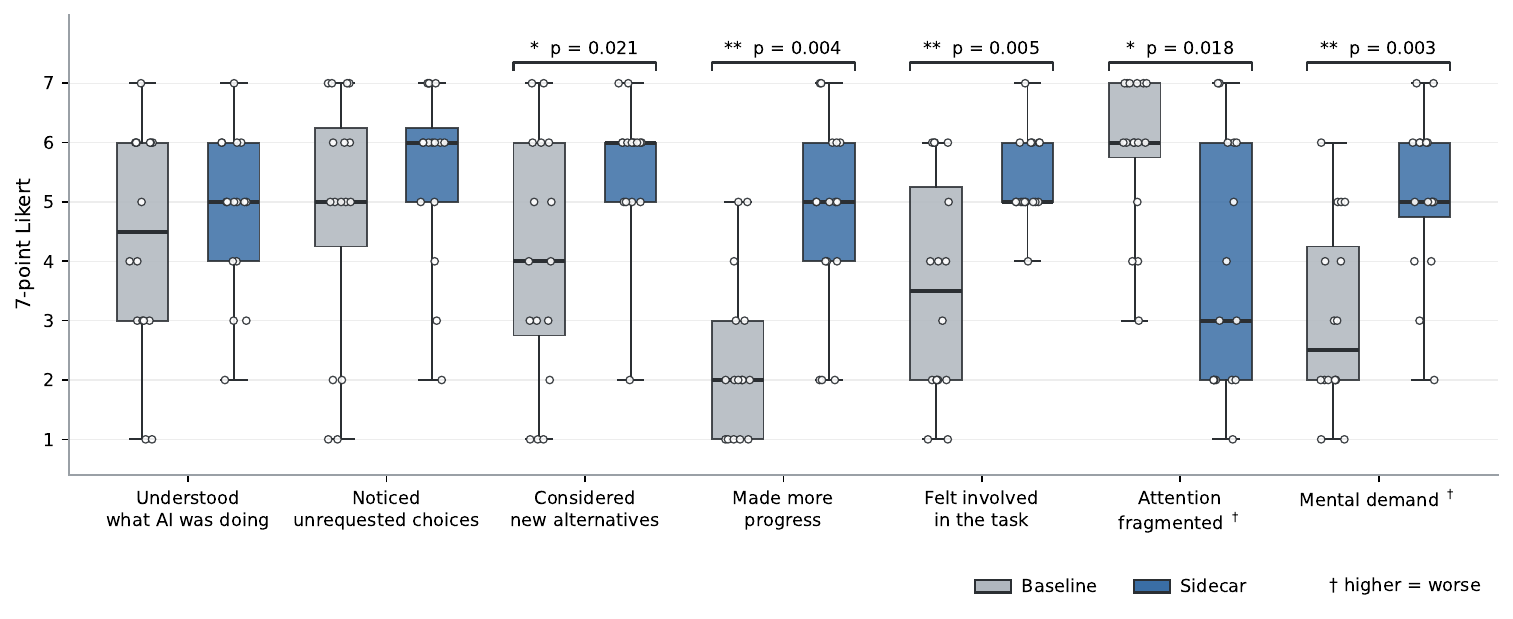}
  \caption{A subset of the self-reported ratings by condition on a 7-point Likert
    scale ($N=16$). Boxes span the interquartile range, the line
    marks the median, whiskers cover the full range, and each
    point is one participant. All items are shown as presented to
    participants, with no reverse-coding; $\dagger$ marks the two items
    where a \emph{higher} score is the worse outcome. Brackets mark items that
    remain significant after Benjamini--Hochberg FDR correction across
    all 11 survey items (two-sided Wilcoxon signed-rank on paired
    within-participant differences).
    (Full survey questions and data in Appendix, Table~\ref{tab:survey-full}).}
  \label{fig:survey}
\end{figure*}

\subsection{Survey results}
Results from the post-study survey are shown in Figure \ref{fig:survey}.
Participants reported feeling less attention fragmented in \sys{} (\sys{} $\text{Mdn}{=}3.0$, baseline $\text{Mdn}{=}6.0$, $p{=}0.018$, lower is better).
With \sys, participants reported making more progress on the task while the agent worked (5.0 vs.\ 2.0, $p{=}0.004$), feeling more involved in what the AI was doing (5.0 vs.\ 3.5, $p{=}0.005$), with a greater consideration of alternatives that they would not have thought of on their own (6.0 vs.\ 4.0, $p{=}0.021$).
However, they also reported higher mental demand (5.0 vs.\ 2.5, $p{=}0.003$) in \sys{}). 
Full item-level results appear in Appendix Table ~\ref{tab:survey-full}. Below we highlight the main findings from the user study.

\subsection{With \sys{}, participants were more focused and engaged on their task}
\label{sec:focus}

With \sys{}, participants stayed on task and reported feeling more engaged and focused while waiting. 
As noted in the behavioral metrics above, users were generally more likely to stay focused on the main task when using \sys, and less likely for their attention to drift elsewhere. Largely, this was because \sys gave them engaging ways to make progress on the task during the interstitial time.
P3 explained, \textit{``I definitely was more engaged. I had things to interact with. I had like decisions to make. So, that part of it felt like fun and like the [wait] time definitely went by much more quickly.''}
V3 described how \sys{} helped them hone in on one task: \textit{``rather than doing two very different tasks in the same amount of time to some decent degree of success, I kind of can just zone into one work stream and not have to switch out but also use my time effectively [...] I was less annoyed because I was having more fun.''}
V4, who normally starts a generation and then is \textit{``doing five other things,''} said that with \sys{} they were \textit{``on it the whole time.''}

By comparison, the baseline condition left participants scattered across other tasks and having to repeatedly check back on the agent.
When listing out what they had done while waiting during the baseline, ten of sixteen participants noted they were not quite happy with it, despite trying to be productive in other ways.
For example, V2 had looked at messages and shopping: \textit{``it feels like you're productive, but all these add up into stress or noise.''}
Aside from self-interrupting and moving to side tasks, six participants also described the secondary interruption associated with leaving the side task to return if the agent stalled or required user input.
\textit{``there's task switching both away from the main task, but also when I'm on a side task, all of a sudden we have to check on that thing,''} with the result that \textit{``you're constantly checking — your vigilance is everywhere.''} (V2)
In contrast, with \sys{}, participants said they noticed the agent finishing simply because they were on the same page.

Finally, task switching in the baseline was substantially more pronounced in the video task, where the interface mainly shows a spinner and a final clip. In contrast, the agent's accumulating code and running trace already gave participants something worth watching directly. Hence, the extent and nature of the agent trace should be a future consideration when designing interventions like \sys. We discuss how to balance the benefits of these interventions with cognitive load in the Discussion.

\subsection{\sys{}'s read-only design and ambient delivery preserved user initiative, but visual salience could be further reduced}
\label{sec:nonactuation}

Participants appreciated that \sys{} could be used to make progress on the main task without unintentionally modifying the code or video in progress.
Two participants noted that when they normally work with agents, they have to explicitly specify to them that they should not begin implementing or acting on the artifact in progress.
P1, for example, typed explicitly in the main chat \textit{``to not start coding''} when they were still brainstorming, because otherwise it might make changes unprompted, leaving them frustrated.
V3 appreciated that \sys{} \textit{would not generate code or an output versus a secondary chat which just tries to edit your code.''}
Echoing this, P5 explained that they liked how \sys{} felt more like \textit{``an external observer.''}

Knowing \sys{} wouldn't act on the artifact, participants felt free to use it as a space to stash and work through rough, unfinished ideas.
Six participants used the side chat to jot down ideas, draft prompts, or save references that came up during the wait but couldn't be sent to the main agent without interrupting it.
P8 explained the motivation behind this: \textit{``I kind of want to tell it this stuff, but I can't yet because then I interrupt it [\ldots] I need this draft workspace where I can put a bunch of stuff.''}
Supporting this, V8 said \textit{``there's something nice about having a companion for working on a script and not having to switch tabs.''}
In effect, \sys{} became a low-stakes space to think alongside the main agent rather than through it.

Finally, participants appreciated that \sys{}'s suggestions were delivered in a way that was easy to ignore and didn't demand attention. Because the system-surfaced decisions and suggestions appeared in small text directly below the side chat (rather than hidden behind a click or appearing as a popup), it was both easy to view and easy to dismiss without requiring any extra user action.
V5 said they liked \sys{}'s \textit{ambient qualities [\ldots] I didn't feel any pressure to click.''}
When asked if they felt pushed to interact with it, P5 said \textit{``not at all. Only by my own curiosity, but not pushed by the tool itself.''}
P1 similarly felt \textit{``it was not a forced interaction, it was just more engagement with the problem.''}
This ease of ignoring \sys{}'s suggestions is also reflected in the near-zero dismissal rate discussed in \S\ref{sec:whatdid}: participants would often leave a \sys{} suggestion unread or read it and move on, rather than feeling the need to dismiss and remove it from the interface. 

That being said, many participants also described attending more to the \sys{} interface instead of the code in the main chat. It is possible that, despite not requiring any user action to dismiss, any visual change in a side chat, even if subtle, could draw user attention and encourage passive monitoring. We discuss these tradeoffs in Discussion below.

\subsection{By surfacing the main agent's decisions, \sys{} made users aware of their own underspecification}
\label{sec:underspec}

\sys{}'s suggestions and extracted decisions helped participants reflect on the prompts they had given to the main agent and the ways in which they were underspecified.
For the video task, V3 remarked that the system's suggestions helped them realize the expansive design space surrounding the video they could generate and encouraged them to \textit{``think about what was missing in my prompt''} and \textit{``infer how the system might interpret the [original] prompt.''}
Meanwhile, in the prototyping task, the extracted decisions unearthed key, technical details the users neglected to specify in their original prompt.
P2 noted that they helped them to find \textit{``places where I had to make decisions that I didn't think about.''}
Interestingly, in the video condition, users found suggestions useful even when they were not completely relevant to their current thought process.
For instance, when creating restaurant promotion videos using dog characters, V2 felt their first clip was \textit{``not promotional enough,''} and wanted the emphasis moved from the dogs onto the restaurant, while \sys{} surfaced suggestions about the dogs' personality and interaction style. 
\kp{V2 felt that they were off-target, \textit{``that's not the original intent,''} but reading them prompted an idea of their own: \textit{``this kind of jogs my memory [\ldots] particularly I think like the branding shot at the end is probably nice.''} They added a closing shot with the restaurant's logo, a refinement the suggestion never proposed.
V2 explained, \textit{``I wouldn't use the [dog personality] suggestions because they weren't relevant, but they helped me branch off and think about other ideas that were relevant.''}}
In more open-ended tasks like video generation, suggestions retained value even when off-target, since the creative space had no single correct answer and an irrelevant idea could still spark a usable one.

Suggestions could be ``irrelevant'' in multiple ways, not just in reference to the user's current thought process but also in what the user can effectively consider.
For example, P7 is a designer without a coding background, and during the prototyping task, was shown a few genuine gaps and considerations for an algorithm their application used.
However, she did not think this suggestion was relevant, because these decisions sat at a layer she deliberately delegated to the main agent to figure out: \textit{``it was things like I didn't know I needed that. So, okay, an algorithm for choosing. I just wanted a sign-up list with menu suggestions.''}
She wanted her suggestions to be at the design layer, for example, \textit{``where did it [the working agent] get this spacing or this font from?''}

\subsection{Suggestions aligning to users' immediate state were highly valued, but those probing too far ahead increased cognitive load}

\label{sec:scope}

Participants valued that \sys{}'s suggestions streamed in live as the agent worked, which let them catch unwanted decisions early rather than after the fact. P8 was surprised suggestions kept pace with the agent in real time: \textit{``I hadn't expected it to be live update. That's kind of neat that it's watching and commenting. It's almost like someone else is sitting next to me looking at the trajectory in real time with me.''} A key part of this was that suggestions were grounded in the agent's actual work; V1 \textit{``liked when it referenced what the thinking was doing as part of its suggestions,''} and V2 read these grounded suggestions as a signal of the agent's state, calling them \textit{``a nice early signal to figure out whether I need to start crafting another prompt.''}

This grounding in the agent's live work helped six participants discover unwanted agent decisions while the run was still in progress rather than only after it finished. P1 contrasted the two conditions directly: \textit{``[With \sys{}] I could see what game it was building while it was building the game itself. So I realized maybe I want something else and I was able to intervene,''} whereas in the baseline \textit{``the first time I really understood what the agent is doing is when I saw the UI [it had built].''}
P6 similarly valued that the \sys{} was \textit{``surfacing out some of the decisions [\ldots] something that I would either not do or do through manual prompting.''} V2 went further, using a flagged divergence as grounds to interrupt the agent mid-run, noting that \sys{} \textit{``does give you a preview of things that deviated from your original prompt [\ldots] which could help you terminate the task early''} and thereby \textit{``save tokens.''}

Suggestions were most useful when they were directly relevant to the user's current thinking or the next prompt they intended to give, rather than suggestions that probed too far ahead.
\kp{V3's session illustrates the difference within a single wait. On the travel video task, one suggestion---to adjust camera movement speed---missed because it assumed a problem that their video might have, even though V3 could not see the output yet: \textit{``I understand what it's saying and why it's saying it, but my video isn't jerky [for camera movement].''} 
In contrast, another suggestion landed on a video style choice they were already weighing: \textit{``I actually don't like this vignette situation. So that is a useful recommendation [\ldots] I was on the fence about the vignette.''} A third suggestion proposed a street-level shot to complement the skyline footage they had been generating, which V3 accepted outright, \textit{``the street level is a good idea,''} and carried into their next prompt.}
Suggestions that explored too far ahead, beyond what the user was immediately considering, increased cognitive load instead.
P3 described this as feeling like multitasking: \textit{``it felt like we were trying to decide the smaller things at the same time as the big things.''}
V1 drew a clear boundary around how far ahead was comfortable: \textit{``thinking one step ahead is comfortable.''}
But even suggestions that stayed one step ahead could become overwhelming if delivered in rapid succession, since each new suggestion pushed the user to think one step ahead all over again.
V1 described this compounding effect: \textit{``I had run one prompt and it was suggesting some stuff to think about, and so I wanted to follow up on that, and as I was following up on that step ahead it looked like it was going to generate another thing to think about [\ldots] so you almost end up in this loop of thinking ahead on the step that you're thinking ahead on.''}
Together, these accounts suggest that wait-time suggestions should adapt to a user's current capacity for cognitive load, pacing delivery and prioritizing what surfaces first, rather than surfacing every next-step consideration at once.

\section{Discussion}
\label{sec:discussion}

In the following section, we discuss: (1) how we might make wait-time assistance more adaptive to manage cognitive load, (2) how \sys{} and future agent wait-time systems might handle multiple agents working at the same time, and (3) what designing for agent wait-times might look like in the future, when agents are potentially faster and more ubiquitous in our daily lives. We end with limitations to our study and future work.

\subsection{Managing User Cognitive Load While Waiting by Adapting Wait-time Assistance}
\label{sec:disc-adaptivity}
While \sys{} enabled users to hone in and focus on their task during the agent wait-time, it also introduced more cognitive load.
This extra mental work stemmed from deliberating on \sys{}'s suggestions and relating them back to the main agent's work.
Our results point to three dimensions along which the right amount of assistance varied between users.
The first is how far ahead a suggestion reaches beyond the main agent's work: V1 was comfortable one step out, but overwhelmed beyond that.
The second is which layer of the problem it addresses; P7 did not want code details but design-level suggestions instead.
The third is when and how often suggestions were provided; having them appear after every prompt had a few participants feeling like they were in a ``loop'' of thinking ahead.

One way to help reduce this cognitive load is to let users tune the wait-time assistance directly.
Participants might define rules for the wait-time system to follow, to dictate what kind of suggestions to give (e.g. on the code, features or the design of the prototype), how forward-thinking or divergent they should be, and when or how often they should appear.
For instance, a user might dictate a rule that the system should only provide divergent suggestions when they are in more exploratory phases of building a prototype (e.g. creating the first version of a brand new feature), and perhaps no suggestions when they are doing targeted debugging.
These rules should be easy for the user to create and edit as they get accustomed to the wait-time system and use it for a variety of different tasks.

In addition to letting users define their own rules, the wait-time system might also infer these preferences from their behavior.
It might leverage signals like repeatedly opening decision cards while leaving exploration cards unread, or dismissing any suggestion that touches the code and concrete implementation, to suggest rules like ``bias toward extracting decisions from the main agent's work'' and ``give me suggestions only on the design of the prototype''.
Inferring these rules might have its own challenges.
A user who ignores the \sys{} during one demanding wait has not asked the system to show less, and a system that quietly retunes on that basis leaves them no way to say so.
We suspect the practical answer combines the two, with inferred adjustments surfaced as proposals the user can accept or decline rather than applied silently. Either way, the design goal from \sys holds: adaptation should make the side chat signals easier to ignore when the user wants less, not work harder to be noticed.

\subsection{Concurrent Monitoring Rather Than Sequential Task-Switching}
\label{sec:disc-monitoring}

Our behavioral and interview data suggest that leaving an agent run during a micro-wait may be more nuanced than the classical interruption model, where attention transfers from one task to another and the primary cost is resumption upon return \cite{altmann_memory_2002, katidioti_interrupt_2016}. Our participants instead described a state of divided monitoring when waiting for agents: because the agent may finish, stall, or request input at any moment, leaving creates two concurrent monitoring obligations rather than a clean handoff (\S\ref{sec:focus}). 
This distinction has implications for how wait-time interventions are designed and measured. Our data suggests that \sys{} reduced time spent away from the agent's tab (\S\ref{sec:focus}), keeping user attention engaged in a way that made staying more worthwhile than leaving. 

While our study asked participants to work on a single task, real-world use of agents may involve inherently more divided monitoring, with users running and waiting on multiple agents at the same time.
An increasingly common behavior during code generation is to run multiple, concurrent agents completing different tasks.
Four participants in the study described doing this.
To keep track of what these agents are doing, P8 described \textit{``skipping between conversations''} On the one hand, systems like \sys could play a critical role in reducing these context-switching behaviors, by helping users stay in flow on fewer tasks, rather than constantly interleaving between them. On the other hand, two participants noted they sometimes run the same prompt in different tabs to produce a variety of results and check in on each of them. In those cases, where users are purposefully performing concurrent, independent runs focused on the same task, making \sys{}'s design even less obtrusive and keeping it user-initiated will be even more crucial, in order not to further divide the user's attention.
For example, future derivatives of \sys{} might sit atop these multiple agent threads, prioritizing ambient suggestions and extracted decisions for runs it deems require the most attention.
Or, they might serve as a way to help users stay less fragmented, by encouraging them to focus on one or two agent tasks in particular, rather than dividing their attention among more.

\subsection{Designing for Agent Micro-waits as Agents become Faster and More Ubiquitous}
\label{sec:disc-future}
As agentic systems and their underlying models improve, the nature of agent waiting might evolve.
Tasks that take multiple minutes today (e.g. implementing a feature) might be completed in a few seconds in the future. At the same time, highly complex tasks (e.g. conducting a large team project, or conducting an open-ended experiment) may reduce from hours to minutes.
The amount of work per unit of time will increase, and after a few minute wait, users will likely have to spend much more of their time verifying and exploring what the agentic system has created.
In this case, a system like \sys{} could help users get ready for this verification, scanning over the work, prioritizing, and proactively surfacing items they should assess while they wait.
And perhaps with increasing agent capabilities, users might be less inclined to inspect the agent's work, confident that what the system has produced is correct and useful.
\sys{} could help bring participants back into the artifacts they are producing and engage them to think critically about it, instead of leaving most of the decision-making to the model.

In the future, agentic systems could also enter more parts of knowledge work and contexts beyond work, extending agent micro-waits from a desktop-bound experience into more varied settings, like waiting on a phone while cooking or waiting on a trip-planning agent while commuting.
These different circumstances raise open questions: first, the form factor of a passive side-chat may need to adapt when we no longer assume a large, single screen for agent work. %
Second, as agentic systems increasingly blend into daily life rather than being confined to a dedicated work session, waits will compete with higher-stakes attention demands (a conversation, a moving deadline, childcare), meaning systems may need to reason not just about how much cognitive bandwidth a suggestion requires, but the user's broader situational context and mental bandwidth beyond work done on their screen.

\subsection{Limitations and Future Work}
Our evaluation involved 16 participants working on assigned tasks in a single lab session. We chose this format to enable controlled, within-subjects comparison across two domains with counterbalanced conditions. The trade-off is that we cannot yet speak to durability: participants had limited time to develop patterns of use, and five mentioned novelty as a factor in their engagement. Notably, participants had limited motivational investment in the task outcome, which could affect how representative their behavior was compared to if real stakes were involved. For example, with greater intrinsic motivation, users may be more likely to do task-related activities in other agent tabs, rather than switching to unrelated activities in the baseline condition while waiting. At the same time, even during a controlled lab study, participants in the baseline condition switched to irrelevant activities such as shopping and texting during waits, suggesting that the attention-fragmentation problem we observe is genuine and may be even more pronounced in natural work settings. Finally, the short task duration also meant that participants were still learning the \sys{} interface and could feel rushed, which may have contributed to the reported mental workload increase. A natural next step would be a longer-term deployment study where participants use \sys{} with their own tasks over days or weeks, which would clarify whether the engagement and attention allocation patterns we observed persist as the interface becomes familiar and the initial novelty fades.

Our study evaluated \sys{} as an integrated system, which was appropriate for understanding the overall experience of wait-time support but does not isolate which design properties drove which effects: shared context vs. independent context in the side chat; the side chat alone vs. the system-suggested decisions to review and directions to explore; the different task domains. Notably, the domain contrast between less (video) and more (prototyping) legible runs produced qualitatively different accounts of how \sys{} helped, and a study designed around this contrast with a larger sample could clarify when each mechanism operates. 

Finally, participants repeatedly expressed a desire to know how long the wait would last, which had motivated us to explore predicting wait duration from early-run signals (as described in Section \ref{sec:nonintrusive-features}). The predictions did not reach sufficient accuracy for pacing decisions, but we report this as it may be informative for others building similar systems: the branching and backtracking structure of agentic reasoning may make agent waits resistant to the kind of duration forecasting that works well for compile times or batch jobs, and investigating why could help calibrate design assumptions for future wait-time interventions. 

\section{Conclusion}
\label{sec:conclusion}

\mxl{
As agentic systems take on more complex work, users increasingly find themselves in micro-waits that are too long to watch and too short to leave. Through interviews with 12 daily agent users, we found that these waits tend to produce diffuse, unproductive vigilance, with users either watching traces they struggle to parse or switching to unrelated activities and losing focus on the task at hand. We built \sys{}, a user interface that presents as a passive side chat sharing full context with the agent's conversation, augmented with capabilities to help users think through their main task during the wait. In a within-subjects evaluation (N=16) across two task domains, \sys{} reduced off-task switching (35\% → 13\% of wait-time), keeping participants focused on their main task. Its passive, ambient design let participants freely brainstorm and stash unfinished ideas without fear of the system acting on their work-in-progress. \sys{} also surfaced the agent's live decisions, helping participants recognize underspecification in their own prompts. Suggestions worked best when they stayed grounded in the agent's immediate trajectory, whereas suggestions that ran too far ahead or arrived too frequently increased cognitive load. As agents grow more capable and micro-waits become a routine part of knowledge work, designing for the quality of these waiting moments may become as important as designing for the quality of the agent's output itself.
}

\bibliographystyle{ACM-Reference-Format}
\bibliography{sample-base}

\appendix
\clearpage
\twocolumn[%
\section{Appendix}
\subsection{Full Survey Results}
  \centering
  \small
  \captionof{table}{All 11 self-report survey items, verbatim as presented to
    participants. Questions for D. Workload are selected from NASA-TLX
    \cite{HART1988139} and adapted to a 7-point scale. The Physical Demand
    question was discarded due to irrelevance to the task. Medians on a
    1--7 scale, shown as presented with no reverse-coding; $p$ from a
    two-sided Wilcoxon signed-rank test on paired within-participant
    differences ($N=16$). Bold rows remain significant after
    Benjamini--Hochberg FDR correction across all 11 items. Items marked
    $\dagger$ are those where a \emph{higher} score is the worse outcome:
    all five workload scales measure demand, and C2 asks about attention
    drifting away. For these, a downward shift is the favourable result.}
  \label{tab:survey-full}
  \begin{tabular}{@{}llccc@{}}
    \toprule
    & \textbf{Question as presented} & \textbf{Baseline} & \textbf{\sys{}} & \textbf{$p$} \\
    \midrule
    \multicolumn{5}{@{}l}{\textit{A. Understanding the agent}} \\
    A1 & I understood what the AI was doing while I waited. & 4.5 & 5.0 & .268 \\
    A2 & I noticed choices the AI made that I had not asked for. & 5.0 & 6.0 & .242 \\
    \multicolumn{5}{@{}l}{\textit{B. Exploration}} \\
    \textbf{B1} & \textbf{I considered alternatives I would not have thought of on my own.} & \textbf{4.0} & \textbf{6.0} & \textbf{.021} \\
    \textbf{B2} & \textbf{I made more progress on the task while the agent is working.} & \textbf{2.0} & \textbf{5.0} & \textbf{.004} \\
    \multicolumn{5}{@{}l}{\textit{C. Engagement}} \\
    \textbf{C1} & \textbf{I felt involved in what the AI was doing.} & \textbf{3.5} & \textbf{5.0} & \textbf{.005} \\
    \textbf{C2}$^\dagger$ & \textbf{My attention drifted away from the task while waiting.} & \textbf{6.0} & \textbf{3.0} & \textbf{.018} \\
    \multicolumn{5}{@{}l}{\textit{D. Workload}} \\
    \textbf{D1}$^\dagger$ & \textbf{Mental demand} & \textbf{2.5} & \textbf{5.0} & \textbf{.003} \\
    D2$^\dagger$ & Temporal demand & 2.5 & 5.0 & .044 \\
    D3$^\dagger$ & Performance (1 - Perfect, 7 - Failure) & 5.0 & 4.0 & .751 \\
    D4$^\dagger$ & Effort & 3.0 & 4.5 & .032 \\
    D5$^\dagger$ & Frustration & 3.0 & 3.5 & .877 \\
    \bottomrule
  \end{tabular}
]

\end{document}